\documentclass[onecolumn, 11pt, letterpaper]{article}

\usepackage[linesnumbered,ruled]{algorithm2e}
\usepackage{amsmath}\allowdisplaybreaks
\usepackage{amssymb}
\usepackage{amsthm}
\usepackage{color}
\usepackage[margin = 1in]{geometry}
\usepackage{hyperref}
\usepackage{mathtools}

\usepackage{algorithmic}
\usepackage{authblk}
\usepackage{thm-restate}

\newtheorem{corollary}{Corollary}
\newtheorem{definition}{Definition}

\newtheorem{lemma}{Lemma}

\newtheorem{theorem}{Theorem}

\title{Simple Deterministic Approximation for Submodular Multiple Knapsack Problem}
\author[1,2]{Xiaoming Sun}
\author[1,2]{Jialin Zhang}
\author[3]{Zhijie Zhang}
\affil[1]{Institute of Computing Technology, Chinese Academy of Sciences}
\affil[2]{University of Chinese Academy of Sciences}
\affil[3]{School of Mathematics and Statistics, Fuzhou University}
\affil[ ]{sunxiaoming@ict.ac.cn, zhangjialin@ict.ac.cn, zzhang@fzu.edu.cn}

\begin{document}
	
	\maketitle
	
	\begin{abstract}
		Submodular maximization has been a central topic in theoretical computer science and combinatorial optimization over the last decades. Plenty of well-performed approximation algorithms have been designed for the problem over a variety of constraints. In this paper, we consider the submodular multiple knapsack problem (SMKP). In SMKP, the profits of each subset of elements are specified by a monotone submodular function. The goal is to find a feasible packing of elements over multiple bins (knapsacks) to maximize the profit. Recently, Fairstein et al.~[ESA20] proposed a nearly optimal $(1-e^{-1}-\epsilon)$-approximation algorithm for SMKP. Their algorithm is obtained by combining configuration LP, a grouping technique for bin packing, and the continuous greedy algorithm for submodular maximization. As a result, the algorithm is somewhat sophisticated and inherently randomized. In this paper, we present an arguably simple deterministic combinatorial algorithm for SMKP, which achieves a $(1-e^{-1}-\epsilon)$-approximation ratio. Our algorithm is based on very different ideas compared with Fairstein et al.~[ESA20].
	\end{abstract}
	
	\section{Introduction}
	\label{Section: introduction}
	
	The \emph{multiple knapsack problem} (MKP) is defined as follows.
	We are given a set $N$ of $n$ elements and a set $M$ of $m$ bins (knapsacks).
	Each element $u\in N$ has a positive cost $c(u)>0$ and a positive profit $p(u)>0$.
	The cost (profit) of a subset $S\subseteq N$ equals the sum of the costs (profits) of its elements.
	The $j$-th bin in $M$ has a positive budget $B_j>0$ for $1\leq j\leq m$.
	A subset $S\subseteq N$ is feasible if there is a disjoint partition $\{S_j\}_{j=1}^m$ of $S$ such that $c(S_j)\leq B_j$ for $1\leq j \leq m$.
	The goal is to find a feasible set $S$ (and its partition $\{S_j\}_{j=1}^m$) whose profit is maximized.
	It is well-known that the problem admits a PTAS but no FPTAS assuming P $\neq$ NP \cite{Kellerer99, ChekuriK00, Jansen09}.
	
	In this paper, we consider the submodular generalization of the above problem, referred to as the \emph{submodular multiple knapsack problem} (SMKP).
	In SMKP, the profit is in general non-additive and specified by a \emph{monotone submodular} function $f: 2^N \rightarrow \mathbb{R}_+$.
	Here, a set function $f: 2^N \rightarrow \mathbb{R}$ is monotone if $f(S)\leq f(T)$ for any $S\subseteq T$ and submodular if $f(S \cup\{u\}) - f(S) \geq f(T \cup\{u\}) - f(T)$ for any $S \subseteq T$ and $u \not \in T$.
	The goal is again to find a feasible set $S$ which maximizes the profit $f(S)$.
	When $m=1$, the problem reduces to submodular maximization under a knapsack constraint, which enjoys an optimal $(1-e^{-1})$-approximation \cite{Kellerer99, Sviridenko04}.
	
	Submodular functions capture the effect of diminishing returns in the economy and generalize many well-known functions such as coverage functions, cut functions, matroid rank functions, and log determinants.
	By introducing a submodular objective, SMKP falls in the field of submodular maximization, which studies maximization problems with submodular objectives, including maximum coverage problem, maximum cut problem, submodular welfare problem \cite{Vondrak08}, influence maximization \cite{KempeKT03}. The study of submodular maximization has lasted for more than forty years.
	As early as 1978, it was shown that for monotone submodular maximization, a greedy algorithm achieves a $(1 - e^{-1})$-approximation under the cardinality constraint \cite{NemhauserWF78} and a $1 / 2$ approximation under the matroid constraint \cite{FisherNW78}.
	On the other hand, even for the cardinality constraint, the problem does not admit an approximation ratio better than $1 - e^{-1}$ \cite{NemhauserW78}.
	It was a longstanding open question whether the problem admits a $(1 - e^{-1})$-approximation under the matroid constraint.
	In 2008, Vondr{\'{a}}k \cite{Vondrak08} made a big breakthrough and answered this question affirmatively by proposing the so-called \emph{continuous greedy} algorithm.
	Since then, plenty of optimal or well-performed approximation algorithms have been proposed for submodular maximization over a variety of constraints \cite{BuchbinderF16, BuchbinderFNS12, BuchbinderFNS14, EneN16, FeldmanNS11, FilmusW12, GharanV11, LeeMNS09, LeeSV10, VondrakCZ11}.
	
	For SMKP, a nearly optimal $(1-e^{-1}-\epsilon)$-approximation algorithm based on the continuous greedy technique was recently proposed in \cite{FairsteinKNRS20}.
	Their algorithm relies on two key ideas.
	First, they showed that by defining a configuration LP, an SMKP instance whose all bins have the same budget can be reduced to submodular maximization under $2$-dimensional packing constraints (SMPC).
	Second, they developed a grouping technique inspired by \cite{VegaL81} to convert a general SMKP instance to a leveled instance where bins are partitioned into blocks and bins in the same block have the same budget.
	In this way, they are able to reduce a general SMKP instance to an SMPC instance.
	They finally finished their work with a refined analysis of the continuous greedy algorithm for SMPC.
	
	The techniques adopted by \cite{FairsteinKNRS20} and the way to combine them are somewhat sophisticated, which makes their algorithm not easy to understand and implement.
	Besides, the continuous greedy technique involves a sampling process and therefore their algorithm is inherently randomized.
	To the best of our knowledge, no deterministic algorithm was known for SMKP.
	In this paper, we present a simple deterministic combinatorial algorithm for SMKP, which achieves a $(1-e^{-1}-\epsilon)$-approximation ratio.
	
	\begin{theorem}
		For any $\epsilon>0$, there exists a deterministic combinatorial algorithm for SMKP that achieves a $(1-e^{-1}-\epsilon)$-approximation ratio and runs in polynomial time.
	\end{theorem}

	\subsection{Technique Overview}
	
	We start with solving SMKP instances under the \emph{identical case}, where all the bins have the same budget $B$.
	Such instances can be reduced to exponential-size instances of submodular maximization subject to a cardinality constraint.
	Inspired by this observation, we design an algorithm for the identical case by mimicking the greedy algorithm for the cardinality constraint.
	See Section \ref{Section: overview-the identical case} for details.
	
	For any general SMKP instance, we use the grouping technique developed by \cite{FairsteinKNRS20} to convert it to the so-called leveled instance.
	While Fairstein et al.~\cite{FairsteinKNRS20} resorts to the configuration LP to solve the leveled instance, we present a simple $(1-e^{-1}-\epsilon)$-approximation algorithm for it by exploiting its structure and invoking our algorithm for the identical case as a subroutine.
	See Section \ref{Section: overview-the general case} for details.
	
	\subsubsection{The Identical Case}
	\label{Section: overview-the identical case}
	
	Under the identical case, SMKP can be regarded as an exponential-size instance of submodular maximization subject to a cardinality constraint.
	Specifically, let $\mathcal{I}=\{S\subseteq N\mid c(S)\leq B\}$.
	For any $\mathcal{T}\subseteq\mathcal{I}$, define $g(\mathcal{T})=f(\cup_{S\in\mathcal{T}} S)$.
	It is easy to verify that $g$ is a monotone submodular function.
	Then, $\max\{g(\mathcal{T})\mid |\mathcal{T}|\leq m\}$ describes the SMKP instance under the identical case.
	
	Inspired by the above observation, our algorithm packs bins one by one and manages to make each bin pack at least the average marginal value of the optimal solution over $m$ bins.
	In other words, for the $j$-th bin, it aims to find a set $S_j$ such that $f(S_j\mid \cup_{i=1}^{j-1}S_i)\geq \frac{1}{m}f(OPT\mid \cup_{i=1}^{j-1}S_i)$, where $OPT$ denotes the optimal solution.
	This naturally leads to $(1-e^{-1})$ approximation.
	
	We take the first bin as an example and explain that it is possible to find a set $S_1$ such that $f(S_1)\geq\frac{1}{m}f(OPT)$ when $m$ is large enough.
	If $S_1$ is obtained by packing elements in sequence greedily according to their marginal densities, then we can prove
	\[ f(S_1)\geq (1-e^{-c(S_1)/c(OPT)})\cdot f(OPT). \]
	If we further allow $S_1$ to violate the budget constraint by adding one more element, then $c(S_1)\geq B$.
	Together with $c(OPT)\leq mB$, we have
	\[ f(S_1)\geq (1-e^{-1/m})\cdot f(OPT)\approx\frac{1}{m} f(OPT). \]
	
	The story has not ended since the last element added to $S_1$ violates the budget constraint.
	To handle this issue, our algorithm divides elements into large and small elements according to their costs and then packs them in different ways.
	Specifically, an element $u\in N$ is large if $c(u)>\epsilon B$ and small otherwise.
	Our algorithm packs large elements by enumeration since there are polynomial ways to pack them in total.
	It packs small elements greedily as before.
	In this way, the last element added to $S_1$ has a cost less than $\epsilon B$ and there are at most $m$ such elements.
	Thus, all of them can be repacked using additional $\epsilon m$ bins and all $S_j$'s will then become feasible.
	
	In Lemma \ref{Lemma: the identical case}, we show that $f(S_1)\geq\frac{1}{m}f(OPT)$ still holds although we introduce the enumeration step.
	
	\subsubsection{The General Case}
	\label{Section: overview-the general case}
	
	Observe that a general SMKP instance can be reduced to an exponential-size instance of submodular maximization subject to a partition matroid constraint.
	Specifically, let $\mathcal{I}_j=\{S\subseteq N\mid c(S)\leq B_j\}$ be the feasible region for the $j$-th bin and  $\mathcal{I}=\cup_{j=1}^m \mathcal{I}_j$.
	For any $\mathcal{T}\subseteq\mathcal{I}$, define $g(\mathcal{T})=f(\cup_{S\in\mathcal{T}}S)$.
	Then, $\max\{g(\mathcal{T})\mid |\mathcal{T}\cap\mathcal{I}_j|\leq 1,1\leq j\leq m\}$ describes the general SMKP instance.
	Recall that the optimal $(1-e^{-1})$-approximation for the partition matroid constraint is obtained via the continuous greedy algorithm \cite{Vondrak08}.
	Thus, it is not a good idea to solve general SMKP instances directly.
	
	The difficulty in solving general SMKP stems from that the budgets are distinct.
	Therefore, we first consider an ``intermediate'' instance where bins can be partitioned into $r$ blocks $\{M_k\}_{k=1}^r$ such that block $M_k$ contains sufficiently many bins and all of them have the same budget $B_k$.
	Clearly, this instance is slightly more general than the instance under the identical case.
	It can also be reduced to an exponential-size instance of submodular maximization subject to a partition matroid constraint.
	Specifically, let $\mathcal{I}_k=\{S\subseteq N\mid c(S)\leq B_k\}$ for $1\leq k\leq r$ and $\mathcal{I}=\cup_{k=1}^r \mathcal{I}_k$.
	For any $\mathcal{T}\subseteq\mathcal{I}$, define $g(\mathcal{T})=f(\cup_{S\in\mathcal{T}}S)$.
	Then, $\max\{g(\mathcal{T})\mid |\mathcal{T}\cap\mathcal{I}_k|\leq |M_k|,1\leq k\leq r\}$ describes the above SMKP instance.
	
	The above two reductions lead to different constraints $|\mathcal{T}\cap\mathcal{I}_j|\leq 1$ and $|\mathcal{T}\cap\mathcal{I}_k|\leq |M_k|$.
	For convenience, assume that $1/\epsilon$ is an integer, $|M_k|\geq 1/\epsilon$ and $\epsilon|M_k|$ is an integer for all $1\leq k\leq r$.
	Our key observation is that for constraint $\{\mathcal{T}\subseteq \mathcal{I}\mid |\mathcal{T}\cap\mathcal{I}_k|\leq |M_k|,1\leq k\leq r\}$, there is a simple deterministic algorithm that can achieve $(1-e^{-1}-\epsilon)$-approximation.
	The algorithm runs in $1/\epsilon$ iterations.
	In each iteration, block $M_k$ is visited in sequence and the algorithm will pack $\epsilon |M_k|$ bins in $M_k$.
	This forms an SMKP instance under the identical case.
	Thus, we can invoke our algorithm for the identical case to solve it.
	
	Finally, we apply a grouping technique from \cite{FairsteinKNRS20} to convert a general instance to a $t$-leveled instance which has blocks $\{M_k\}_{k=1}^r$ and bins in the same block have the same budget.
	Besides, each of the first $t^2$ blocks contains a single bin, and each of the remaining blocks contains at least $t$.
	This is very similar to the intermediate instance before and it is not difficult to handle the first $t^2$ blocks.

	\subsection{Related Work}
	\label{Subsection: related work}
	
	MKP has been fully studied previously.
	Kellerer \cite{Kellerer99} proposed the first PTAS for the identical case of the problem.
	Soon after, Chekuri and Khanna \cite{ChekuriK00} proposed a PTAS for the general case.
	The result was later improved to an EPTAS by Jansen \cite{Jansen09}.
	On the other hand, it is easy to see that the problem does not admit an FPTAS even for the case of $m = 2$ bins unless P $=$ NP \cite{ChekuriK00}.
	
	SMKP contains submodular maximization subject to a knapsack constraint as a special case.
	For this problem, there is an optimal $(1 - e^{-1})$-approximation algorithm that runs in $O(n^5)$ time \cite{Kellerer99, Sviridenko04}.
	Later, a fast algorithm was proposed in \cite{BadanidiyuruV14} that achieves a $(1 - e^{-1} -\epsilon)$-approximation ratio and runs in $n^2(\log n / \epsilon)^{O(1 / \epsilon^8)}$ time\footnote{As pointed out by \cite{Yoshida19,EneN19}, the result in \cite{BadanidiyuruV14} has some issues.}.
	This was recently improved in \cite{EneN19} by a new algorithm that runs in $(1 / \epsilon)^{O(1 / \epsilon^4)} n \log^2 n$ time.
	The last two algorithms are impractical due to their high dependence on $1/\epsilon$.
	Very recently, a $(1 - e^{-1})$-approximation algorithm was proposed in \cite{KulikSS21,FeldmanNS23}, which runs in $O(n^4)$ time.
	This algorithm can be further accelerated to achieve $(1 - e^{-1} -\epsilon)$-approximation in $\widetilde{O}(n^3/\epsilon)$ time.
	
	To the best of our knowledge, SMKP was first considered in Feldman's Ph.\,D thesis \cite{Feldman13}.
	Feldman proposed a polynomial time $(1 / 9 - o(1))$-approximation algorithm and a pseudo-polynomial time $1/4$ approximation algorithm for the general case of SMKP.
	For the identical case, he improved the results to a polynomial time $((e - 1) / (3e - 1) - o(1))\approx 0.24$ approximation algorithm and a pseudo-polynomial time $(1 - e^{-1} - o(1))$-approximation algorithm.
	These algorithms are based on the \emph{continuous greedy} technique and \emph{contension resolution schemes} \cite{VondrakCZ11}, and hence involve randomness inherently.
	Recently, Fairstein et al.~\cite{FairsteinKNRS20} proposed a polynomial time randomized $(1-e^{-1}-\epsilon)$-approximation algorithm for general SMKP.

	\subsection{Organization}
	
	In Section \ref{Section: preliminaries}, we first formulate SMKP and introduce some notations.
	Then, we present a greedy algorithm that packs elements greedily according to their marginal densities.
	In Section \ref{Section: the identical case}, we present a $(1-e^{-1}-\epsilon)$-approximation algorithm for SMKP under the identical case, assuming the number of bins $m\geq 1/(4\epsilon^3)$.
	In Section \ref{Section: the general case}, we present a $(1-e^{-1}-\epsilon)$-approximation algorithm for general SMKP.
	We conclude the paper and list some open problems in Section \ref{Section: conclusion}.

	\section{Preliminaries}
	\label{Section: preliminaries}

	An instance of the \emph{submodular multiple knapsack problem} (SMKP) is defined as follows.
	We are given a set $N$ of $n$ elements and a set $M$ of $m$ bins (knapsacks).
	Each element $u\in N$ has a positive cost $c(u)>0$.
	A subset $S\subseteq N$ of elements has a cost $c(S)=\sum_{u\in S} c(u)$.
	The $j$-th bin in $M$ has a positive budget $B_j>0$ for $1\leq j\leq m$.
	A subset $S\subseteq N$ is feasible for the problem if there is a disjoint partition $\{S_j\}_{j=1}^m$ of $S$ such that $c(S_j)\leq B_j$ for $1\leq j \leq m$.
	The profit of each subset $S\subseteq N$ of elements is specified by a \emph{normalized}, \emph{monotone} and \emph{submodular} function $f: 2^N \rightarrow \mathbb{R}_+$.
	For a non-negative set function $f: 2^N \rightarrow \mathbb{R}_+$, it is called \emph{normalized} if $f(\emptyset)=0$, \emph{monotone} if $f(S) \leq f(T)$ for any $S \subseteq T$, and \emph{submodular} if $f(S \cup\{u\}) - f(S) \geq f(T \cup \{u\}) - f(T)$ for any $S \subseteq T$ and $u \not \in T$.
	The goal is to find a feasible set $S$ (and its partition $\{S_j\}_{j=1}^m$) such that the profit $f(S)$ (or $f(\cup_{j=1}^m S_j)$) is maximized.
	
	An SMKP instance is specified by $(N,M,\{B_j\}_{j\in M},f,c)$.
	Throughout this paper, we use $OPT$ to denote the optimal solution of an SMKP instance.
	Let $S + u$ be a shorthand for $S \cup \{u\}$.
	For the objective function $f$, we also use $f(u \mid S)$ and $f(T \mid S)$ to denote the marginal values $f(S + u) - f(S)$ and $f(S \cup T) - f(S)$, respectively.
	$f$ is accessed via a value oracle that returns $f(S)$ when set $S\subseteq N$ is queried.
	The query complexity of any algorithm for SMKP should be polynomial in the size of the problem.
	
	\subsection{The Greedy Algorithm}
	\label{Subsection: the greedy algorithm}
	
	We first present a greedy algorithm, which is depicted as Algorithm \ref{Algorithm: greedy}.
	It serves as a cornerstone for other algorithms in this paper.
	It returns a (possibly infeasible) set with a $(1 - 1 / e)$ approximation ratio.
	It packs elements one by one greedily, according to their \emph{densities}, namely the ratios of their marginal values to their costs.
	The process continues provided there exists some bin whose budget has not been exhausted yet.
	As a side effect, each bin may pack one more element whose addition exceeds the budget of that bin.
	For convenience, we refer to this element as a \emph{reserved element}.
	Nonetheless, we show that the set returned by Algorithm \ref{Algorithm: greedy} has a large profit.
	
	\begin{algorithm}[tb]
		\SetAlgoLined
		\KwIn{elements $N$, budgets $\{B_j\}_{j=1}^m$, profit $f$, cost $c$.}
		
		$S_j=\emptyset$ for $1\leq j\leq m$ and $S=\cup_{j=1}^m S_j$.
		
		\While{$N \setminus S \neq \emptyset$ and there exists $1\leq j \leq m$ such that $c(S_j) < B_j$}{
			$u^* = \arg \max_{u \in N \setminus S} f(u \mid S) / c(u)$.
			
			$S_j = S_j + u^*$ and $S = S + u^*$.
		}
		\Return $S=\cup_{j=1}^m S_j$.
		\caption{\textsc{Greedy}}
		\label{Algorithm: greedy}
	\end{algorithm}
	
	
	\begin{lemma}
		\label{Lemma: greedy}
		Let $S$ be the set returned by Algorithm \ref{Algorithm: greedy}.
		For any set $X \subseteq N$, we have
		\[ f(S) \geq \left( 1 - e^{-c(S)/c(X)} \right) \cdot f(X). \]
	\end{lemma}
	
	\begin{proof}
		If $c(S)<\sum_{j=1}^{m} B_j$, there is some $j$ such that $c(S_j)<B_j$.
		It means that  Algorithm \ref{Algorithm: greedy} ended with $S=N$.
		Thus, the lemma follows by monotonicity.
		
		Now consider the case where $c(S) \geq \sum_{j=1}^{m} B_j$.
		Assume that $S = \{ u_1, u_2, \ldots, u_{\ell} \}$, and for $0\leq i\leq \ell$, $S^i = \{u_1, u_2, \ldots u_i\}$ denotes the first $i$ elements packed by Algorithm \ref{Algorithm: greedy}.
		Then, by the greedy rule,
		\[ \frac{f(u_i \mid S^{i - 1})}{c(u_i)} \geq \frac{f(x \mid S^{i - 1})}{c(x)}, \forall\, x \in X \setminus S^{i - 1}. \]
		By moving $c(x)$ to the left and summing over $x\in X \setminus S^{i - 1}$,
		\[ c(X \setminus S^{i - 1})\cdot\frac{f(u_i \mid S^{i - 1})}{c(u_i)}  \geq \sum_{x\in X \setminus S^{i - 1}} f(x \mid S^{i - 1}) \geq f(X \setminus S^{i - 1} \mid S^{i - 1}). \]
		The last inequality holds since $f$ is submodular.
		This gives us
		\begin{equation}
			\frac{f(S^i) - f(S^{i - 1})}{c(u_i)} \geq \frac{f(X \setminus S^{i - 1} \mid S^{i - 1})}{c(X \setminus S^{i - 1})} \geq \frac{f(X) - f(S^{i - 1})}{c(X)}. \label{eq: greedy}
		\end{equation}
		The last inequality holds since $f$ is monotone and $c(X \setminus S^{i - 1}) \leq c(X)$.
		
		Next, we assume that $f(X) > f(S^{\ell})$, since otherwise the lemma already holds.
		Under this assumption, it must hold that $c(u_i) < c(X)$, since otherwise inequality \eqref{eq: greedy} implies that $f(X) \leq f(S^i) \leq f(S^{\ell})$.
		A contradiction!
		Now we can rearrange inequality \eqref{eq: greedy} and obtain that
		\[ f(X) - f(S^i) \leq \left(1 - \frac{c(u_i)}{c(X)}\right)(f(X) - f(S^{i - 1})). \]
		By expanding the recurrence, we have
		\[ f(X) - f(S^i) \leq \prod_{j = 1}^{i} \left(1 - \frac{c(u_j)}{c(X)}\right) \cdot f(X) \leq \prod_{j = 1}^{i} e^{- \frac{c(u_j)}{c(X)}} \cdot f(X) = e^{-\frac{c(S^i)}{c(X)}} \cdot f(X). \]
		The second inequality holds due to $e^x \geq 1 + x$.
		Hence we have
		\[ f(S^i) \geq \left( 1 - e^{-c(S^i)/c(X)} \right) \cdot f(X). \]
		The lemma follows by plugging $i=\ell$ into it.
	\end{proof}
	
	The above lemma immediately leads to the following corollary.
	
	
	\begin{corollary}
		\label{Corollary: greedy}
		The set $S$ returned by Algorithm \ref{Algorithm: greedy} satisfies $f(S) \geq (1 - e^{-1})\cdot f(OPT)$.
	\end{corollary}
	
	\begin{proof}
		If $c(S)<\sum_{j=1}^{m} B_j$, there is some $j$ such that $c(S_j)<B_j$.
		It means that  Algorithm \ref{Algorithm: greedy} ended with $S=N$.
		Thus, the corollary follows by monotonicity.
		If $c(S) \geq \sum_{j=1}^{m} B_j$, then $c(S) \geq c(OPT)$.
		The corollary follows from Lemma~\ref{Lemma: greedy}.
	\end{proof}

	\section{The identical Case}
	\label{Section: the identical case}
	
	In this section, we present a deterministic $(1-e^{-1}-\epsilon)$ approximation algorithm for SMKP under the identical case, where all bins have the same budget.
	Our algorithm is depicted as Algorithm \ref{Algorithm: the identical case} and works when $m\geq 1/(4\epsilon^3)$.
	It packs bins one by one and manages to make each bin pack at least the average marginal value of the optimal solution over $m$ bins.
	In other words, for the $j$-th bin, it aims to find a set $S_j$ such that $f(S_j\mid \cup_{i=1}^{j-1}S_i)\geq \frac{1}{m}f(OPT\mid \cup_{i=1}^{j-1}S_i)$.
	This naturally leads to $(1-e^{-1})$ approximation.
	For this purpose, Algorithm \ref{Algorithm: the identical case} divides elements into \emph{large} and \emph{small} elements according to their costs.
	Given input $\epsilon$, an element $u \in N$ is large if $c(u) > \epsilon B$ and small otherwise.
	Let $N_{\ell} = \{u \in N \mid c(u) > \epsilon B\}$ be the set of large elements and $N_s = N \setminus N_{\ell}$.
	For the $j$-th bin, Algorithm \ref{Algorithm: the identical case} first enumerates all feasible subsets of large elements.
	Then, for every such subset, Algorithm \ref{Algorithm: greedy} is invoked over small elements to augment it.
	Finally, the one with the maximum marginal value is assigned to $S_j$.
	
	Due to the call of Algorithm \ref{Algorithm: greedy}, $S_j$ might contain a reserved element, which is the last added into $S_j$ and violates the budget.
	To remedy this issue, Algorithm \ref{Algorithm: the identical case} divides the bins into two classes: the first $(1 - \epsilon)m$ bins are called \emph{working bins} and the last $\epsilon m$ bins are called \emph{reserved bins}.
	The procedure described above only proceeds with the working bins.
	After that, Algorithm \ref{Algorithm: the identical case} repacks all reserved elements into the reserved bins.
	We will show that in this way, Algorithm \ref{Algorithm: the identical case} produces a feasible solution and the loss of the profit is little even if it does not use the reserved bins to pack new elements.
	
	We now give an analysis of Algorithm \ref{Algorithm: the identical case}.
	For $1\leq j\leq (1-\epsilon)m$, let $S_j$ be defined as in line \ref{line: Sj} of Algorithm \ref{Algorithm: the identical case} and $T_j=\cup_{i=1}^j S_i$.
	We first show that Algorithm \ref{Algorithm: the identical case} returns a feasible solution.
	
	\begin{lemma}
		Algorithm \ref{Algorithm: the identical case} produces a feasible solution.
	\end{lemma}
	
	\begin{proof}
		For $1\leq j \leq (1-\epsilon)m$, observe that each $S_j$ contains at most one reserved element due to the call of Algorithm \ref{Algorithm: greedy}.
		By repacking those reserved elements into the reserved bins, each $S_j$ becomes feasible.
		Besides, the cost of each reserved element is at most $\epsilon B$ since it is a small element.
		Thus, a reserved bin can pack at least $1/\epsilon$ reserved elements.
		Then, $\epsilon m$ reserved bins can pack $m>(1-\epsilon)m$ reserved elements without exceeding their budgets.
		Therefore, Algorithm \ref{Algorithm: the identical case} produces a feasible solution.
	\end{proof}
	
	Next, we present Lemma \ref{Lemma: the identical case} for Algorithm \ref{Algorithm: the identical case}.
	
	\begin{algorithm}[tb]
		\SetAlgoLined
		\KwIn{elements $N$, budget $B$, number of bins $m$, profit $f$, cost $c$, constant $\epsilon > 0$.}
		
		Let the first $(1-\epsilon)m$ bins be \emph{working bins} and the last $\epsilon m$ bins be \emph{reserved bins}. 
		
		Define $N_{\ell} = \{u \in N \mid c(u) > \epsilon B\}$ and let $N_s = N \setminus N_{\ell}$.
		
		$S_j=\emptyset$ for $1\leq j\leq m$ and $T=\cup_{j=1}^m S_j$.
		
		\For{$j=1$ to $(1-\epsilon)m$}{
			\ForEach{subset $E\subseteq N_{\ell}$ such that $c(E)\leq B$}{
				$G_E =$ \textsc{Greedy}$(N_s, B-c(E), f(\cdot \mid T\cup E), c(\cdot))$.
			}
			$S_j=\arg\max_E f(E\cup G_E\mid T)$ and $T=\cup_{j=1}^m S_j$. \label{line: Sj}
			
		}
		
		Repack the reserved elements in $T$ into the reserved bins.
		
		\Return $T=\cup_{j=1}^m S_j$.
		\caption{\textsc{Identical-case}}
		\label{Algorithm: the identical case}
	\end{algorithm}
	
	\begin{lemma}
		\label{Lemma: the identical case}
		Assume that $m \geq 1 / (4 \epsilon^3)$.
		For every $1\leq j\leq (1-\epsilon)m$,
		\[ f(S_j\mid T_{j-1}) \geq \frac{1 - 2\epsilon}{m}\cdot f(OPT\mid T_{j-1}). \]
	\end{lemma}
	
	\begin{proof}
		For the sake of description, we define $g(\cdot)=f(\cdot\mid T_{j-1})$ and the lemma becomes $g(S_j) \geq \frac{1 - 2\epsilon}{m}\cdot g(OPT)$.
		Let $OPT_{\ell}=OPT\cap N_{\ell}$ and $OPT_{s}=OPT\setminus OPT_{\ell}$.
		We prove the lemma by case analysis, according to the cost and density of $OPT_s$.
		
		\textbf{Case 1:} $c(OPT_s)\geq\epsilon mB$, namely $OPT_s$ has a large cost.
		Let $OPT_{\ell}=\cup_{j=1}^m OPT_{\ell,j}$ and $OPT_{s}=\cup_{j=1}^m OPT_{s,j}$, where $OPT_{\ell,j}$ and $OPT_{s,j}$ are the large and small elements packed in the $j$-th bin, respectively.
		For each $1\leq j\leq m$, since $c(OPT_{\ell,j})\leq B$, $OPT_{\ell,j}$ will be enumerated during the \textbf{foreach} loop.
		Let $G_j$ be the output of \textsc{Greedy} (Algorithm \ref{Algorithm: greedy}) starting from $OPT_{\ell,j}$.
		We will show that one of $OPT_{\ell,j}\cup G_j$ satisfies the lemma.
		
		If $c(G_j)< B-c(OPT_{\ell,j})$, it means that Algorithm \ref{Algorithm: greedy} ended with $G_j=OPT_s$ and therefore $g(G_j\mid OPT_{\ell,j})=g(OPT_s\mid OPT_{\ell,j})$.
		If $c(G_j)\geq B-c(OPT_{\ell,j})$, then $c(G_j)\geq c(OPT_{s,j})$.
		By Lemma \ref{Lemma: greedy},
		\begin{align*}
			g(G_j\mid OPT_{\ell,j}) &\geq \left(1-e^{-c(G_j)/c(OPT_s)}\right)\cdot g(OPT_s\mid OPT_{\ell,j}) \\
			&\geq \left(1-e^{-c(OPT_{s,j})/c(OPT_s)}\right)\cdot g(OPT_s\mid OPT_{\ell,j}) \\
			&\geq \left(\frac{c(OPT_{s,j})}{c(OPT_s)} - \frac{c(OPT_{s,j})^2}{2 \cdot c(OPT_s)^2}\right) \cdot g(OPT_s\mid OPT_{\ell,j}) \\
			&\geq \left(\frac{c(OPT_{s,j})}{c(OPT_s)} - \frac{1}{2\epsilon^2m^2}\right) \cdot g(OPT_s\mid OPT_{\ell,j}) \\
			&\geq \left(\frac{c(OPT_{s,j})}{c(OPT_s)} - \frac{2\epsilon}{m}\right) \cdot g(OPT_s\mid OPT_{\ell,j}).
		\end{align*}
		The third inequality holds since $1 - e^{-x} \geq x - x^2 / 2$ for $x \geq 0$.
		The fourth inequality holds since $c(OPT_{s,j})/c(OPT_s) \leq 1/(\epsilon m)$.
		The last inequality holds since $m \geq 1 / (4 \epsilon^3)$.
		
		By adding $g(OPT_{\ell,j})$ on both sides of the last inequality and summing over $j$,
		\begin{align*}
			\sum_{j=1}^{m} g(OPT_{\ell,j}\cup G_j) &\geq \sum_{j=1}^{m}\left(\frac{c(OPT_{s,j})}{c(OPT_s)} - \frac{2\epsilon}{m}\right) \cdot g(OPT_s\mid OPT_{\ell,j})+\sum_{j=1}^{m}g(OPT_{\ell,j}) \\
			&\geq \sum_{j=1}^{m}\left(\frac{c(OPT_{s,j})}{c(OPT_s)} - \frac{2\epsilon}{m}\right) \cdot g(OPT_s\mid OPT_{\ell})+g(OPT_{\ell}) \\
			&=(1-2\epsilon)\cdot g(OPT_s\mid OPT_{\ell})+g(OPT_{\ell}) \\
			&\geq (1-2\epsilon)\cdot g(OPT).
		\end{align*}
		Hence, the maximum of $OPT_{\ell,j}\cup G_j$ satisfies the lemma and so does $S_j$.
		
		\textbf{Case 2:} $g(OPT_s) \geq (1 - e^{-B / c(OPT_s)})^{-1} \cdot \frac{g(OPT)}{m}$, namely the density of $OPT_s$ is large.
		Consider one of the iterations of \textbf{foreach} loop where $E=\emptyset$.
		Note that it is augmented by $G_{\emptyset}$ via \textsc{Greedy} (Algorithm \ref{Algorithm: greedy}).
		If $c(G_{\emptyset})<B$, it means that Algorithm \ref{Algorithm: greedy} ended with $G_{\emptyset}=OPT_s$.
		Then,
		\[ g(G_{\emptyset})= g(OPT_s)\geq (1 - e^{-B / c(OPT_s)})^{-1} \cdot \frac{g(OPT)}{m}\geq \frac{g(OPT)}{m}. \] 
		If $c(G_{\emptyset})\geq B$, by Lemma \ref{Lemma: greedy},
		\[ g(G_{\emptyset}) \geq \left(1 - e^{-B / c(OPT_s)}\right) \cdot g(OPT_s) \geq \frac{g(OPT)}{m}. \]
		This implies that $G_{\emptyset}$ satisfies the lemma and so does $S_j$.
		
		\textbf{Case 3:} $c(OPT_s) < \epsilon mB$ and $g(OPT_s) < \left(1 - e^{-B / c(OPT_s)}\right)^{-1} \cdot \frac{g(OPT)}{m}$, namely both the cost and density of $OPT_s$ are small.
		We show that $OPT_s$ only contributes a negligible value in $OPT$:
		\begin{align*}
			g(OPT_s) &< (1 - e^{-1 / \epsilon m})^{-1} \cdot \frac{g(OPT)}{m} \leq \left(\frac{1}{\epsilon m} - \frac{1}{2\epsilon^2 m^2}\right)^{-1} \frac{g(OPT)}{m} \\
			&\leq \left(\frac{1}{2 \epsilon m}\right)^{-1} \frac{g(OPT)}{m} = 2\epsilon\cdot g(OPT).
		\end{align*}
		The first inequality holds since $(1 - e^{-B/x})^{-1}$ is monotone increasing.
		The second holds since $1 - e^{-x} \geq x - x^2 / 2$ for $x \geq 0$.
		The third holds as long as $m \geq 1 / \epsilon$.
		Hence, by submodularity,
		\[ g(OPT_{\ell}) \geq g(OPT) - g(OPT_s) \geq (1 - 2\epsilon) \cdot g(OPT), \]
		and 
		\[ \frac{1}{m}\sum_{j=1}^{m} g(OPT_{\ell,j})\geq \frac{1}{m} \cdot g(OPT_{\ell}) \geq \frac{1 - 2\epsilon}{m} \cdot g(OPT). \]
		This implies that the maximum of $OPT_{\ell,j}$ satisfies the lemma and so does $S_j$.
	\end{proof}
	
	By expanding the recurrence in Lemma \ref{Lemma: the identical case}, we have
	\begin{lemma}
		\label{Lemma: bound for Tj}
		Assume that $m\geq 1/(4\epsilon^3)$.
		For every $1\leq j\leq (1-\epsilon)m$,
		\[ f(T_j)\geq (1-e^{-j(1-2\epsilon)/m})\cdot f(OPT). \]
	\end{lemma}
	\begin{proof}
		By Lemma \ref{Lemma: the identical case}, for $1\leq j \leq (1-\epsilon)m$,
		\[ f(S_j\mid T_{j - 1}) \geq \frac{1 - 2 \epsilon}{m}\cdot f(OPT\mid T_{j-1}). \]
		By monotonicity of $f$,
		\[ f(T_j) - f(T_{j - 1}) \geq \frac{1 - 2 \epsilon}{m}\cdot (f(OPT) - f(T_{j - 1})). \]
		By rearranging the above inequality,
		\[ \left(1 - \frac{1 - 2 \epsilon}{m}\right) (f(OPT) - f(T_{j - 1})) \geq f(OPT) - f(T_j). \]
		By expanding the recurrence,
		\[ f(OPT) - f(T_j) \leq \left(1 - \frac{1 - 2 \epsilon}{m}\right)^j f(OPT)\leq e^{-j(1-2\epsilon)/m}\cdot f(OPT). \]
		The last inequality holds since $e^{-x}\geq 1-x$.
		Thus, we have
		\[ f(T_j)\geq (1-e^{-j(1-2\epsilon)/m})\cdot f(OPT). \]
	\end{proof}
	
	We now provide a theoretical guarantee for Algorithm \ref{Algorithm: the identical case}.
	
	\begin{theorem}
		\label{Theorem: guarantee of the identical case algorithm}
		When $m \geq 1 / (4\epsilon^3)$, Algorithm \ref{Algorithm: the identical case} achieves a $(1 - e^{-1} - O(\epsilon))$ approximation ratio and uses $O(mn^{3+1/\epsilon})$ queries.
	\end{theorem}
	
	\begin{proof}
		For the approximation ratio, by plugging $j=(1-\epsilon)m$ into Lemma \ref{Lemma: bound for Tj},
		\[ f(T_{(1-\epsilon)m})\geq (1 - e^{-(1-\epsilon)(1 - 2\epsilon)})\cdot f(OPT). \]
		
		For the query complexity, observe that during the \textbf{foreach} loop, the number of subsets $E\subseteq N_{\ell}$ such that $c(E)\leq B$ is at most
		\[ \sum_{i=0}^{1/\epsilon}\binom{n}{i}=O(n^{1/\epsilon+1}). \]
		Since each $E$ is augmented via \textsc{Greedy}, which uses $O(n^2)$ queries, the \textbf{foreach} loop uses $O(n^{1/\epsilon+3})$ in total.
		Then, Algorithm \ref{Algorithm: the identical case} overall uses $O(mn^{3+1/\epsilon})$ queries.
		
	\end{proof}
	

	%
	%
	
	\section{The General Case}
	\label{Section: the general case}
	
	In this section, we present a deterministic $(1-e^{-1}-\epsilon)$ approximation algorithm for solving general SMKP instances.
	A key difficulty is that the budgets of bins are distinct, which makes our technique for the identical case inapplicable.
	In Section \ref{Section: reshape the instance}, we introduce a grouping technique from \cite{FairsteinKNRS20}, which reshapes any SMKP instance such that bins can be partitioned into \emph{blocks} and almost every block contains sufficiently many bins with the same budget.
	Next, in Section \ref{Section: the final algorithm}, we show how one can design a nearly optimal algorithm for such instances.
	
	\subsection{Reshape the Instance}
	\label{Section: reshape the instance}
	
	We first introduce a grouping technique from \cite{FairsteinKNRS20} to reshape any SMKP instance as follows.
	
	\begin{definition}
		A subset of bins $M'\subseteq M$ is called a block if for any $i,j\in M'$, $B_i=B_j$.
	\end{definition}
	
	\begin{definition}
		For any $t\in\mathbb{N}_+$, a partition $\{M_k\}_{k=1}^r$ of bins $M$ is $t$-leveled if for every $1\leq k \leq r$, $M_k$ is a block and $|M_k|=t^{\lfloor (k-1)/t^2\rfloor}$.
	\end{definition}
	To gain some intuition, note that for every $1\leq k\leq t^2$, block $M_k$ contains a single bin, and for every $t^2<k\leq 2t^2$, block $M_k$ contains $t$ bins, etc.
	It follows that except for the first $t^2$ blocks, each of the remaining blocks contains at least $t$ bins with the same budget.
	
	\begin{lemma}[\cite{FairsteinKNRS20}]
		\label{Lemma: reshape the instance}
		There is a polynomial-time algorithm, referred to as \textsc{Block}, that takes a set of bins $M$, budgets $\{B_j\}_{j\in M}$ and a parameter $t\in\mathbb{N}_+$ as input, and returns a new set of bins $\widetilde{M}\subseteq M$, budgets $\{\widetilde{B}_j\}_{j\in \widetilde{M}}$ and a $t$-leveled partition $\{\widetilde{M}_k\}_{k=1}^r$ of bins $\widetilde{M}$ such that
		\begin{itemize}
			\item For every $j\in \widetilde{M}$, $\widetilde{B}_j\leq B_j$.
			\item For any SMKP instance $(N,M,\{B_j\}_{j\in M},f,c)$ and a feasible solution $\{S_j\}_{j\in M}$ for it, there exists a feasible solution $\{\widetilde{S}_j\}_{j\in\widetilde{M}}$ for instance $(N,\widetilde{M},\{\widetilde{B}_j\}_{j\in \widetilde{M}},f,c)$ such that $f(\cup_{j\in \widetilde{M}}\,\widetilde{S}_j)\geq \left(1-\frac{1}{t}\right) f(\cup_{j\in M} S_j)$ and $\cup_{j\in \widetilde{M}}\,\widetilde{S}_j\subseteq \cup_{j\in M} S_j$.
		\end{itemize}
	\end{lemma}
	
	The instance $(N,\widetilde{M},\{\widetilde{B}_j\}_{j\in \widetilde{M}},f,c)$ is called $t$-\emph{leveled}.
	Lemma \ref{Lemma: reshape the instance} tells us that any feasible solution for it is also feasible for the original instance $(N,M,\{B_j\}_{j\in M},f,c)$, and an optimal solution for it causes a small loss in the profit.
	
	%
	
	\subsection{The Final Algorithm}
	\label{Section: the final algorithm}
	
	Now, we explain how one can design a nearly optimal algorithm for a $t$-leveled SMKP instance with bins $\widetilde{M}$, budgets $\{\widetilde{B}_j\}_{j\in \widetilde{M}}$ and a $t$-leveled partition $\{\widetilde{M}_k\}_{k=1}^r$ of $\widetilde{M}$.
	
	For $t^2<k\leq r$, block $\widetilde{M}_k$ contains $|\widetilde{M}_k|\geq t$ bins with the same budget $\widetilde{B}_k$.
	The problem restricted to each block $\widetilde{M}_k$ can be regarded as an SMKP instance under the identical case.
	Thus, a natural idea is to pack each block $\widetilde{M}_k$ in sequence by invoking Algorithm \ref{Algorithm: the identical case}.
	However, we fail to get an optimal approximation via this procedure.
	Instead, we develop a technique that is inspired by \cite{BadanidiyuruV14}.
	We run $1/\epsilon$ iterations in total (assume that $1/\epsilon$ is an integer).
	In each iteration, we pack each block $\widetilde{M}_k$ in sequence but only pack $\epsilon |\widetilde{M}_k|$ bins (assume that $\epsilon |\widetilde{M}_k|$ is an integer).
	This forms an instance under the identical case with $\epsilon |\widetilde{M}_k|$ bins and therefore we can invoke Algorithm \ref{Algorithm: the identical case} to solve it.
	
	For $1\leq k\leq t^2$, block $\widetilde{M}_k$ contains a single bin with budget $\widetilde{B}_k$.
	Basically, we can use \textsc{Greedy} to pack elements.
	Likewise, we do not use the full budget at a time.
	Instead, we also run $1/\epsilon$ iterations.
	In each iteration, we pack elements using budget $(\epsilon-\epsilon^2) \widetilde{B}_k$.
	To avoid exceeding the budget, we only pack small elements $u$ satisfying $c(u)\leq \epsilon^2 \widetilde{B}_k$.
	To ensure this, we need to enumerate large-valued and large-cost elements in this bin.
	The overall procedure is depicted as Algorithm \ref{Algorithm: the final algorithm for SMKP}.
	
	%
	%
	%
	%
	%
	%
	%
	%
	%
	%
	%
	%

	\begin{algorithm}[tb]
		\SetAlgoLined
		\KwIn{elements $N$, bins $M$, budgets $\{B_j\}_{j\in M}$, profit $f$, cost $c$, constant $\epsilon > 0$.}
		
		Let $s=1/(16\epsilon^9)$ and $t=1/(4\epsilon^3)$.
		
		Let $\mathcal{C}=\emptyset$.
		
		$(\widetilde{M},\{\widetilde{B}_j\}_{j\in\widetilde{M}},\{\widetilde{M}_k\}_{k=0}^r)=$ \textsc{Block}$(M,\{B_j\}_{j\in M},t)$.
		
		\ForEach($\backslash\backslash\,E_j=\emptyset$ for $j>t^2$){feasible solution $\{E_j\}_{j=1}^{t^2}$ such that $|\cup_{j=1}^m E_j|\leq s$}{
			Let $E=\cup_{j=1}^{t^2} E_j$.
			
			Let $S_j=E_j$ for $1\leq j\leq t^2$ and $S_j=\emptyset$ for $t^2<j\leq k$.
			
			\For{$i=1$ to $1/\epsilon$}{
				\For($\backslash\backslash$ handle blocks one by one){$k=1$ to $r$}{
					\uIf($\backslash\backslash$ each block contains a single bin){$k\leq t^2$}{
						Let $D = \{u \in N \setminus E \mid f(u \mid E) > \frac{1}{s}\cdot f(E)\}$.
						
						Let $L_{k}=\{u\in N\setminus E \mid c(u)>\epsilon^2 (\widetilde{B}_k-c(E_k))\}$.
						
						$R_{k}=$ \textsc{Greedy}($N\setminus(E\cup D\cup L_k)$, $(\epsilon-\epsilon^2) (\widetilde{B}_k-c(E_k))$, $f(\cdot\mid \cup_{j=1}^m S_j)$, $c(\cdot)$). \label{line: greedy}
						
						$S_{k+1}=S_{k+1}\cup R_{k}$.
					}
					\Else($\backslash\backslash$ each block contains $\geq t$ bins){
						$\{R_j\}_{j\in \widetilde{M}_k}=$ \textsc{Identical-case}$(N\setminus E, \widetilde{B}_k, \epsilon |\widetilde{M}_k|, f(\cdot\mid \cup_{j=1}^m S_j), c(\cdot), \epsilon)$. \label{line: identical case}
						
						$S_{j}= S_{j}\cup R_{j}$ for $j\in \widetilde{M}_k$.
					}
				}
				
					%
					%
					%
				%
					%
					%
			}
			$\mathcal{C}=\mathcal{C}\cup\{\{S_j\}_{j=1}^m \}$.
		}
		\Return $\arg\max \{f(\cup_{j=1}^m S_j)\mid \{S_j\}_{j=1}^m\in\mathcal{C}\}$.
		\caption{The Final Algorithm for SMKP}
		\label{Algorithm: the final algorithm for SMKP}
	\end{algorithm}
	
	\begin{theorem}
		Algorithm \ref{Algorithm: the final algorithm for SMKP} achieves a $1-e^{-1}-O(\epsilon)$ approximation ratio and uses a polynomial number of queries.
	\end{theorem}
	
	\begin{proof}
		Let $\widetilde{OPT}=\cup_{j=1}^m \widetilde{OPT}_j$ be the optimal solution of the SMKP instance with bins $\widetilde{M}$ and budgets $\{\widetilde{B}_j\}_{j\in\widetilde{M}}$.
		Let $OPT'=\cup_{j=1}^{t^2} \widetilde{OPT}_j$.
		Order elements in $OPT'$ greedily according to their marginal values such that $o_1=\arg\max_{o\in OPT'} f(o)$, $o_2=\arg\max_{o\in OPT' \setminus \{o_1\}} f(o \mid o_1)$, etc.
		Denote by $E$ the first $s$ elements in $OPT'$ (if $|OPT'|<s$, then $E=OPT'$).
		Let $E_j=E\cap \widetilde{OPT}_j$ for $1\leq j\leq t^2$.
		Then, $\{E_j\}_{j=1}^{t^2}$ will be enumerated during the \textbf{foreach} loop.
		In the following, we focus on this particular set.
		
		Let $D=\{u\in N\mid f(u\mid E)> \frac{1}{s}\cdot f(E) \}$.
		Since $E$ is the first $s$ elements in $OPT'$, we have $f(o\mid E)\leq \frac{1}{s}\cdot f(E)$ for any $o\in OPT'\setminus E$.
		Thus,  $D\cap (OPT'\setminus E)=\emptyset$ and therefore $OPT'\setminus E$ will not be excluded from the execution of \textsc{Greedy} over $N\setminus(E\cup D)$.
		Besides, $\{\widetilde{OPT}_j\setminus E_j\}_{j=1}^{t^2}$ is a feasible solution given budgets $\{\widetilde{B}_j-c(E_j)\}_{j=1}^{t^2}$.
		
		For $1\leq j\leq t^2$, let $L_{j}=\{u\in N\setminus E\mid c(u)>\epsilon^2 (\widetilde{B}_j-c(E_j))\}$.
		Define $OPT^*$ as follows.
		For $1\leq j\leq t^2$, $OPT^*_j=\widetilde{OPT}_j\setminus L_j$.
		For $j>t^2$, $OPT^*_j=\widetilde{OPT}_j$.
		Then, $OPT^*=\cup_{j=1}^m OPT^*_j$.
		We have
		\begin{align*}
			f(OPT^*\mid E) &= f((\cup_{j=1}^{t^2}\widetilde{OPT}_j\setminus L_j)\cup(\cup_{j=t^2+1}^{m} \widetilde{OPT}_j)\mid E) \\
			&\geq f(\cup_{j=1}^{m} \widetilde{OPT}_j\mid E)-f(\cup_{j=1}^{t^2} \widetilde{OPT}_j\cap L_j\mid E) \\
			&\geq f(\widetilde{OPT}\mid E)-\sum_{j=1}^{t^2}\sum_{u\in (\widetilde{OPT}_j\setminus E)\cap L_j}f(u\mid E) \\
			&\geq f(\widetilde{OPT}\mid E) -\frac{t^2}{\epsilon^2 s}\cdot f(E) \\
			&= f(\widetilde{OPT}\mid E) -\epsilon \cdot f(E).
		\end{align*}
		The first two inequalities are due to submodularity.
		The third inequality holds since by definition of $L_j$, $\widetilde{OPT}_j\setminus E$ contains at most $1/\epsilon^2$ elements in $L_j$, and $f(u\mid E)\leq \frac{1}{s} f(E)$ due to $D\cap (\widetilde{OPT}_j\setminus E)=\emptyset$.
		The last equality follows from the choices of $t$ and $s$.
		This implies that invoking \textsc{Greedy} over $N\setminus(E\cup D\cup L_j)$ for $1\leq j\leq t^2$ only incurs little loss in the profit.
		
		Now we are prepared to provide a theoretical bound for Algorithm \ref{Algorithm: the final algorithm for SMKP}.
		Let $g(\cdot)=f(\cdot\mid E)$.
		For $1\leq i\leq 1/\epsilon$ and $1\leq k \leq r$, let $R_{ik}$ be the set returned in line \ref{line: greedy} if $k\leq t^2$ and $R_{ik}=\cup_{j\in\widetilde{M}_k}R_j$ otherwise, where $\{R_j\}_{j\in\widetilde{M}_k}$ is the set returned in line \ref{line: identical case}.
		Then, the $k$-th block $\widetilde{M}_k$ packs $\cup_{i=1}^{1/\epsilon} R_{ik}$ by the end of Algorithm \ref{Algorithm: the final algorithm for SMKP}.
		Define $T_0=\emptyset$ and $T_i=T_{i-1}\cup(\cup_{k=1}^r R_{ik})$ for $1\leq i\leq 1/\epsilon$.
		
		For $1\leq i\leq 1/\epsilon$ and $1\leq k\leq t^2$, by Lemma \ref{Lemma: greedy},
		\begin{align*}
			g(R_{ik}\mid T_{i-1}\cup(\cup_{k'=1}^{k-1} R_{ik'}) ) &\geq (1-e^{-(\epsilon-\epsilon^2)})\cdot g(OPT^*_k\mid T_{i-1}\cup (\cup_{k'=1}^{k-1} R_{ik'})) \\
			&\geq (\epsilon-2\epsilon^2)\cdot g(OPT^*_k\mid T_i).
		\end{align*}
		The last inequality holds due to $1-e^{-x}\geq x-x^2/2$ for $x\geq 0$ and submodularity.
		
		For $1\leq i\leq 1/\epsilon$ and $t^2< k\leq r$, by Lemma \ref{Lemma: bound for Tj},
		\begin{align*}
			g(R_{ik}\mid T_{i-1}\cup(\cup_{k'=1}^{k-1} R_{ik'}) ) &\geq (1-e^{-\epsilon(1-2\epsilon)})\cdot g(OPT^*_k\mid T_{i-1}\cup (\cup_{k'=1}^{k-1} R_{ik'})) \\
			&\geq (\epsilon-3\epsilon^2)\cdot g(OPT^*_k\mid T_i).
		\end{align*}
		Again, the last inequality holds due to $1-e^{-x}\geq x-x^2/2$ for $x\geq 0$ and submodularity.
		
		Summing up over $1\leq k\leq r$, we have
		\begin{align*}
			g(T_i)-g(T_{i-1}) &= \sum_{k=1}^{r} g(R_{ik}\mid T_{i-1}\cup(\cup_{k'=1}^{k-1} R_{ik'}) ) \\
			&\geq\sum_{k=1}^{r}(\epsilon-3\epsilon^2)\cdot g(OPT^*_k\mid T_i) \\
			&\geq(\epsilon-3\epsilon^2)\cdot g(OPT^*\mid T_i) \\
			&\geq(\epsilon-3\epsilon^2)\cdot (g(OPT^*)-g(T_i)).
		\end{align*}
		The last two inequalities are due to submodularity and monotonicity, respectively.
		By adding $g(OPT^*)$ to both sides and move $g(T_i)$ to the right in the above inequality,
		\[ g(OPT^*)-g(T_{i-1})\geq(1+\epsilon-3\epsilon^2)(g(OPT^*)-g(T_i)). \]
		This leads to
		\[ g(OPT^*)-g(T_i)\leq \frac{1}{(1+\epsilon-3\epsilon^2)^i}\cdot g(OPT^*). \]
		Hence, by plugging $i=1/\epsilon$,
		\begin{align*}
			g(T_{1/\epsilon}) &\geq\left(1-\frac{1}{(1+\epsilon-3\epsilon^2)^{1/\epsilon}}\right) \cdot g(OPT^*)= \left(1-e^{-\frac{1}{\epsilon}\ln(1+\epsilon-3\epsilon^2)}\right) \cdot g(OPT^*) \\
			&\geq (1-e^{-\frac{1}{\epsilon}(\epsilon-3\epsilon^2-(\epsilon-3\epsilon^2)^2/2)})\cdot g(OPT^*)\geq(1-e^{-1}-O(\epsilon))\cdot g(OPT^*).
		\end{align*}
		The second inequality holds since $\ln(1+x)\geq x-x^2/2$ for $x>0$.
		Finally, recall that $g(\cdot)=f(\cdot\mid E)$, we have
		\begin{align*}
			f(T_{1/\epsilon}) &=f(E)+f(T_{1/\epsilon}\mid E)\geq f(E)+(1-e^{-1}-O(\epsilon))\cdot f(OPT^*\mid E) \\
			&\geq f(E)+(1-e^{-1}-O(\epsilon))\cdot (f(\widetilde{OPT}\mid E) -\epsilon f(E)) \\
			&\geq(1-e^{-1}-O(\epsilon))\cdot f(OPT).
		\end{align*}
		
	\end{proof}
	
	\section{Conclusion}
	\label{Section: conclusion}
	
	In this paper, we present a deterministic $(1-e^{-1}-\epsilon)$-approximation algorithm for SMKP.
	Our algorithm is inspired by the viewpoint regarding SMKP instances as exponential-size instances of submodular maximization subject to a cardinality or partition matroid constraint.
	Thus our algorithm is conceptually much simpler than that of Fairstein et al.~\cite{FairsteinKNRS20}.
	
	As pointed out by \cite{FairsteinKNRS20}, it remains open to remove the loss of $\epsilon$ in the approximation ratio.
	As a first step, we present a $(1-e^{-1})$-approximation algorithm for SMKP when the number of bins $m$ is constant in the appendix.
	Recently, a randomized $0.385$-approximation algorithm for \emph{non-monotone} SMKP was proposed in \cite{FairsteinKS21}.
	It is an interesting question to design deterministic algorithms for this problem.

	
	\bibliographystyle{plain}
	\bibliography{SMKP-arXiv}
	
	\newpage
	\appendix
	
	\section{Constant Number of Bins}
	\label{Section: constant number of bins}
	
	In this section, we present a deterministic $(1 - 1 / e)$ approximation algorithm for SMKP when the number of bins $m$ is a constant.
	We already know that Algorithm \ref{Algorithm: greedy} returns a set with a $(1 - 1 /e)$ approximation ratio.
	However, this set might be infeasible, with a reserved element in each bin.
	To resolve this issue, observe that if those reserved elements have small profits, we are able to discard them directly without losing too much.
	However, Algorithm \ref{Algorithm: greedy} itself can not guarantee this property.
	In light of this,
	Algorithm \ref{Algorithm: constant number of bins} manages to first pack large-value elements in some optimal solution by the enumeration technique, and then pack elements of small value by the greedy algorithm.
	In doing so, it ensures that the values of the reserved elements are small and therefore can be safely discarded.
	
	\begin{algorithm}[tb]
		\SetAlgoLined
		\KwIn{elements $N$, budgets $\{B_j\}_{j=1}^m$, profit $f$, cost $c$, threshold $\delta$.}
		
		\ForEach{feasible solution $\{E_j\}_{j=1}^m$ such that $|\cup_{j=1}^m E_j|\leq\lceil 1/\delta\rceil$}{
				Let $E=\cup_{j=1}^m E_j$ and $D = \{u \in N \setminus E \mid f(u \mid E) > \delta\cdot f(E)\}$.
				
				$G'_E =$ \textsc{Greedy}$(N \setminus (E \cup D), \{B_j-c(E_j)\}_{j=1}^m, f(\cdot \mid E), c(\cdot))$.
				
				Let $R_E \subseteq G'_E$ be the reserved elements in $G'_E$ and $G_E = G'_E \setminus R_E$.
				
				Let $S_E = E \cup G_E$.
			}
		\Return $S=\arg\max\{S_E\mid \mbox{feasible } E \mbox{ that is enumerated}\}$
		
		%
		
		%
				%
						%
						%
						%
		
		\caption{Constant Number of Bins}
		\label{Algorithm: constant number of bins}
	\end{algorithm}
	
	
	\begin{theorem}
		\label{Theorem: constant number of bins}
		If we set $\delta=1/em$, Algorithm \ref{Algorithm: constant number of bins} achieves a $(1 - 1 / e)$ approximation ratio and uses $O((mn)^{em+4})$ queries, which is polynomial when $m$ is a constant.
	\end{theorem}
	
	\begin{proof}
		Assume w.l.o.g.~that $|OPT| > \lceil 1 / \delta \rceil$, since otherwise $OPT$ will be enumerated in the enumeration step.
		We order elements in $OPT$ greedily according to their marginal values, i.e.~$o_1=\arg\max_{o\in OPT} f(o)$, $o_2=\arg\max_{o\in OPT \setminus \{o_1\}} f(o \mid o_1)$, etc.
		In the enumeration step, the solution $E = (E_1, \ldots, E_m)$ must be visited such that $E$ contains exactly the first $\lceil 1 / \delta \rceil$ elements in $OPT$ and these elements are packed in the same way as in $OPT$.
		In the following analysis, we focus on this solution and show that $S_E = E \cup G_E$ achieves the desired ratio.
		Since the algorithm returns the solution with the maximum value, this completes the proof.
		
		We claim that $f(o \mid E) \leq \delta f(E)$ for any $o\in OPT\setminus E$.
		Let $OPT_i$ be the first $i$ elements in $OPT$.
		Then for $j \leq i$ and any $o \not \in OPT_i$, $f(o_j \mid OPT_{j-1}) \geq f(o \mid OPT_{j-1}) \geq f(o \mid OPT_i)$.
		Summing up from $j = 1$ to $i$, we have $f(OPT_i) \geq i \cdot f(o \mid OPT_i)$.
		By plugging $i = \lceil 1 / \delta \rceil$, $f(E) \geq \lceil 1 / \delta \rceil f(o \mid E) \geq \frac{1}{\delta}\cdot f(o \mid E)$.
		
		The above claim implies that $D \cap (OPT \setminus E)=\emptyset$.
		As a result, elements in $OPT \setminus E$ will not be excluded from the execution of the greedy algorithm.
		Besides, since elements in $E$ are packed in the same way as in $OPT$, $OPT \setminus E$ is a feasible (indeed optimal) solution while invoking the greedy algorithm with budgets $\{B_j - c(E_j)\}_{j=1}^m$.
		
		By Corollary \ref{Corollary: greedy}, the set $G'_E$ returned by \textsc{Greedy} satisfies
		\[ f(G'_E \mid E) \geq (1 - 1 / e) f(OPT \setminus E \mid E). \]
		Since $G_E$ is obtained from $G'_E$ by discarding at most $m$ reserved elements in $R_E$, by the submodularity of $f$,
		\begin{align*}
				f(G_E \mid E) &\geq f(G'_E \mid E) - f(R_E \mid E) \\
				&\geq (1 - 1 / e) f(OPT \setminus E \mid E) - \sum_{u \in R_E} f(u \mid E) \\
				&\geq (1 - 1 / e) f(OPT) - (1 - 1 / e) f(E) - \delta m f(E).
			\end{align*}
		Hence we have
		\[ f(E \cup G_E) \geq (1 - 1 / e) f(OPT) + (1 / e - \delta m) f(E) \geq (1 - 1 / e) f(OPT). \]
		The last inequality holds since $1 / \delta \geq em$.
		
		Finally, the number of feasible solutions $\{E_j\}_{j=1}^m$ with $|\cup_{j=1}^m E_j| \leq \lceil 1 / \delta \rceil$ is at most
		\[ \sum_{i=0}^{\lceil1/\delta\rceil}{n\choose i}m^i=O((mn)^{\lceil1 / \delta\rceil+1})=O((mn)^{em + 2}). \]
		Starting from every feasible $E$, Algorithm \ref{Algorithm: constant number of bins} invokes the greedy algorithm which costs $O(n^2)$ queries. Thus, it uses $O((mn)^{em+4})$ queries in total.
		Since $m$ is a constant, it runs in polynomial time.
	\end{proof}

\end{document}